\documentclass[prc, amsfonts, amssymb, amsmath,preprintnumbers,reprint, showkeys, nofootinbib, superscriptaddress]{revtex4-1}
\usepackage[english]{babel}
\usepackage[utf8]{inputenc}
\usepackage[colorinlistoftodos, color=green!40, prependcaption]{todonotes}
\usepackage{amsthm}
\usepackage{mathtools}
\usepackage{physics}
\usepackage{xcolor}
\usepackage{graphicx}
\usepackage[left=23mm,right=13mm,top=35mm,columnsep=15pt]{geometry} 
\usepackage{adjustbox}
\usepackage{placeins}
\usepackage[T1]{fontenc}
\usepackage{lipsum}
\usepackage{csquotes}
\usepackage{lineno}
\newcommand{\alphas}{\ensuremath{\alpha_{s}}}
\newcommand\pT{\ensuremath{p_T}}

\newcommand\pTjet{\ensuremath{p_T^{\rm jet}}}

\newcommand\GeVsquare{\ensuremath{\mathrm{GeV}^{2}}}
\newcommand\nn{\nonumber}

\newcommand\diff{{\rm d}}

\usepackage[normalem]{ulem}

\begin{document}

\title{Jet-based measurements of Sivers and Collins asymmetries \\ at the future Electron-Ion Collider}

\author{Miguel Arratia}
\email{miguel.arratia@ucr.edu}
\affiliation{Department of Physics and Astronomy, University of California, Riverside, CA 92521, USA}
\affiliation{Thomas Jefferson National Accelerator Facility, Newport News, VA 23606, USA}

\author{Zhong-Bo Kang}
\email{zkang@physics.ucla.edu}
\affiliation{Department of Physics and Astronomy, University of California, Los Angeles, CA 90095, USA}
\affiliation{Mani L. Bhaumik Institute for Theoretical Physics, University of California, Los Angeles, CA 90095, USA}
\affiliation{Center for Frontiers in Nuclear Science, Stony Brook University, Stony Brook, NY 11794, USA}

\author{Alexei Prokudin}
\email{prokudin@jlab.org}
\affiliation{Division of Science, Penn State University Berks, Reading, PA 19610, USA}
\affiliation{Thomas Jefferson National Accelerator Facility, Newport News, VA 23606, USA}

\author{Felix Ringer}
\email{fmringer@lbl.gov}
\affiliation{Nuclear Science Division, Lawrence Berkeley National Laboratory, Berkeley, CA 94720, USA}
\affiliation{Physics Department, University of California, Berkeley, CA 94720, USA}

\date{\today}

\begin{abstract}
We present predictions and projections for hadron-in-jet measurements and electron-jet azimuthal correlations at the future Electron-Ion Collider (EIC). These observables directly probe the three-dimensional (3D) structure of hadrons, in particular, the quark transversity and Sivers parton distributions and the Collins fragmentation functions. We explore the feasibility of these experimental measurements by detector simulations and discuss detector requirements. We conclude that jet observables have the potential to enhance the 3D imaging EIC program.
\end{abstract}
\keywords{Electron-Ion Collider, Jets, Three-Dimensional Hadron Structure, Sivers asymmetry, Collins asymmetry, JLAB-PHY-20-3223}

\preprint{JLAB-PHY-20-3223}
\maketitle

\section{Introduction} \label{sec:outline}

Jets, collimated sprays of particles,  observed in high-energy particle collisions, offer a unique opportunity to study  quantum chromodynamics (QCD). Measurements of jets at the Large Hadron Collider (LHC) have triggered the development of new theoretical and experimental techniques for detailed studies of QCD~\cite{Larkoski:2017jix}. 

Jet observables can probe the three-dimensional (3D) hadron structure encoded in transverse-momentum-dependent parton-distribution functions (TMD PDFs) and fragmentation functions (TMD FFs). For example, Higgs-plus-jet production at the LHC gives access to the gluon TMD PDF~\cite{Boer:2014lka}, while the hadron transverse-momentum distribution inside jets probes TMD FFs~\cite{Aad:2011sc,Aaij:2019ctd,Kang:2019ahe}. Recently, jet production in deep-inelastic scattering (DIS) regime was proposed as a key channel for TMD studies~\cite{Gutierrez-Reyes:2018qez,Gutierrez-Reyes:2019vbx,Liu:2018trl,Kang:2020xyq}.  Jets produced in polarized proton-proton collisions probe the Sivers~\cite{Boer:2003tx,Abelev:2007ii,Bomhof:2007su}, transversity and Collins TMDs~\cite{Adamczyk:2017wld,Kang:2017btw,DAlesio:2017bvu}.   

The advent of the Electron-Ion Collider~\cite{Accardi:2012qut} with its high luminosity and polarized beams will unlock the full potential of jets as tools for TMD studies. Measurements of jets in DIS make it possible to control parton kinematics in a way that is not feasible in hadronic collisions. The measurement of jets in DIS will complement semi-inclusive DIS (SIDIS) observables. Generally, jets are better proxies of parton-level dynamics, and they allow for a clean separation of the target and current-fragmentation regions, which is difficult for hadrons~\cite{Boglione:2016bph,Boglione:2019nwk,Aschenauer:2019kzf,Arratia:2019vju}. The measurement of both SIDIS and jet observables is critical to test universality aspects of TMDs within QCD factorization and probe TMD-evolution effects.

The TMD factorization for SIDIS involves a convolution of TMD PDFs and TMD FFs. The observed hadron transverse momentum, $\vec{P}_{hT} = z \vec{k}_T +\vec{p}_T$, receives contributions from both TMD PDFs ($k_T$) and TMD FFs ($p_T$). Here $z$ is the longitudinal-momentum fraction of the quark momentum carried by the hadron. Therefore, it is not possible to separately extract TMD PDFs and TMD FFs in SIDIS alone. Instead, one has to rely on additional processes, such as $e^+e^-$ annihilation and Drell-Yan production. 
One of the major advantages of jet measurements is that they separate TMD PDFs from TMD FFs. 

\begin{figure}[t]
\centering
\vspace*{-1.2cm}
\includegraphics[width=0.4\textwidth]{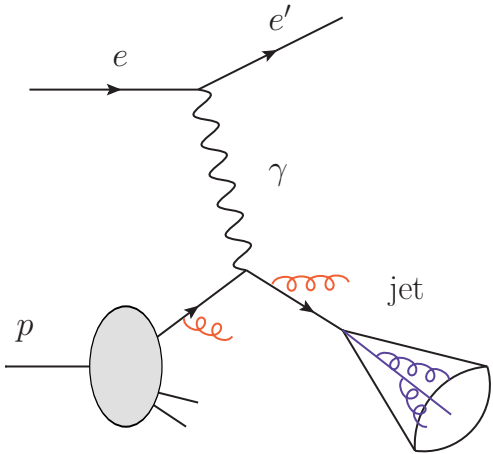}
\vspace*{-1.2cm}
\caption{Illustration of the neutral current DIS process where a jet is recoiling the final-state electron in the laboratory frame.~\label{fig:NC_DIS}}
\end{figure}

We consider back-to-back electron-jet production in the laboratory frame (see Figure~\ref{fig:NC_DIS}),
\begin{align}
e+p(\vec s_T)\to e+({\rm jet}(\vec q_T)\,h(z_h, \vec j_T))+X\,,
\label{eq:main-process}
\end{align}
where $\vec{q}_T$ is the imbalance  of the transverse momentum of the final-state electron and jet (rather than the jet transverse momentum itself), and additionally the transverse momentum $\vec{j}_T$ of hadrons inside the jet with respect to the jet axis. Here $\vec s_T$ is the transverse spin vector of the incoming proton. The imbalance $\vec q_T$ is only sensitive to TMD PDFs~\cite{Liu:2018trl,Buffing:2018ggv}, while the $\vec j_T$ is sensitive to TMD FFs alone~\cite{Bain:2016rrv,Kang:2017glf,Kang:2019ahe}. As a consequence, independent constraints of both TMD PDFs and TMD FFs can be achieved through a single measurement of jets in DIS. In addition, the process considered in this work can be related to similar cross sections accessible in proton-proton collisions, see for example Refs.~\cite{Sun:2015doa,Sun:2018icb}.

An alternative way to isolate TMD PDFs was proposed in Refs.~\cite{Gutierrez-Reyes:2018qez,Gutierrez-Reyes:2019vbx}, using the Breit frame. In this case, the final state TMD dynamics contribute, but they can be evaluated purely perturbatively and TMD FFs are not required. 

The use of jets at the EIC will further benefit from the developments at the LHC and RHIC such as jet reclustering with a recoil-free jet axis~\cite{Bertolini:2013iqa,Chien:2020hzh} or jet-grooming techniques~\cite{Dasgupta:2013ihk,Larkoski:2014wba} which can test QCD-factorization~\cite{Gutierrez-Reyes:2019msa}, probe TMD evolution~\cite{Makris:2017arq,Cal:2019gxa} and explore novel hadronization effects~\cite{Neill:2016vbi,Neill:2018wtk}.

At RHIC, the first and only polarized proton-proton collider, the STAR collaboration pioneered the use of jets for TMD studies. In particular, measurements of the azimuthal asymmetries of hadrons with respect to the jet axis in transversely polarized proton-proton collisions $(pp^{\uparrow})$ probe the Collins fragmentation functions and the collinear transversity distribution~\cite{Adamczyk:2017wld}. As shown in~\cite{Yuan:2007nd,Kang:2017btw}, the in-jet dynamics or the final-state TMD FFs is decoupled from the purely collinear initial state, which provides direct constraints for the Collins TMD FF. The STAR data agree with theoretical predictions~\cite{Kang:2017btw} which rely on transversity functions extracted from SIDIS and $e^+e^-$ data. The current precision of STAR measurements, however, does not allow for clear tests~\cite{Kang:2017btw} of TMD-evolution effects; future measurements will help in this respect~\cite{Aschenauer:2016our}. 

Previous work on EIC projections of TMD measurements focused mainly on SIDIS observables involving either single hadrons or di-hadrons as well as charmed mesons to access gluon TMDs~\cite{Anselmino:2011ay,Accardi:2012qut,Matevosyan:2015gwa}. The feasibility of a gluon Sivers function measurement with di-jets from photon-gluon fusion process at the EIC was explored in Ref.~\cite{Zheng:2018awe}.

In this work, we consider the process in Eq.~\eqref{eq:main-process} in transversely polarized electron-proton collisions, which probes the quark Sivers function, the transversity distribution, and the Collins fragmentation function. We present the first prediction of hadron-in-jet asymmetries at the EIC. In addition, we estimate the precision of EIC data and compare to the uncertainties of predicted asymmetries. We use parametrized detector simulations to estimate resolution effects and discuss requirements for the EIC detectors.

This paper is organized as follows. First we introduce the perturbative QCD framework in Section~\ref{sec:theory}. We then describe the \textsc{Pythia8} simulations used for this study in Section~\ref{sec:simulation}. In Section~\ref{sec:projections}, we present predictions and statistical projections. We discuss jet kinematics as well as detector requirements in Section~\ref{sec:detector} and we conclude in Section~\ref{sec:conclusions}.

\section{Perturbative QCD framework~\label{sec:theory}}

We consider both Sivers and Collins asymmetries at the EIC which can be accessed through jet-based measurements. At the parton level, we consider the leading-order DIS process $eq\to eq$. The cross section is  differential in the electron rapidity $y_e$ and the transverse momentum $p_{T}^e$, which is defined relative to the beam direction in the laboratory frame. The leading-order cross section can be written as
\begin{equation}
    \frac{\diff \sigma}{\diff y_e\,\diff^2 \vec p^{\; e}_T}=\sigma_0\sum_q e_q^2\, f_q(x,\mu) \,,
\end{equation}
where the scale is chosen at the order of the hard scale of the process $\mu\sim p_T^e=|\vec p_T^{\, e}|$. The prefactor $\sigma_0$ is given by
\begin{equation}\label{eq:sigma0}
    \sigma_0=\frac{\alpha\alphas}{s Q^2}\frac{2(\hat s^2+\hat u^2)}{\hat t^2} \,.
\end{equation}
The Bjorken $x$ variable can be expressed as:
\begin{equation}
    x=\frac{p_T^e e^{y_e}}{\sqrt{s} -p_T^e e^{-y_e}} \,.
\end{equation}
Also the partonic Mandelstam variables in Eq.~(\ref{eq:sigma0}) can be expressed in terms of the kinematical variables of the electron and the center-of-mass energy. We have
\begin{align}
    \hat s=&\, xs \,,
    \\
    \hat t=&\, -Q^2 =  -\sqrt{s}\,p_T^e e^{y_e}=-x \sqrt{s}\, p_T^{\rm jet} e^{-y_{\rm jet}} \,,\label{eq:that}
    \\
    \hat u=&\, -x \sqrt{s}\,p_T^e e^{-y_e}= -\sqrt{s}\, p_T^{\rm jet} e^{y_{\rm jet}}\,,
\end{align}
where $p_T^{\rm jet}$ and $y_{\rm jet}$ denote the jet transverse momentum and rapidity, respectively. From Eq.~(\ref{eq:that}), we see that a cut on $Q^2$ translates to an allowed range of the observed $p_T^e,\, y_e$. The event inelasticity $y$, can be written as
\begin{equation}
    y=1-\frac{p_T^e}{\sqrt{s}}e^{-y_e} \,,
\end{equation}
which is an important quantity for experimental considerations as discussed below.

\subsection{The Sivers asymmetry and electron-jet decorrelation}

To access TMD dynamics we study back-to-back electron-jet production, 
\begin{align}
    e+p(\vec s_T)\to e+{\rm jet}(\vec q_T)+X\,,
\end{align}
where we require a small imbalance $q_T=|\vec p_T^{\, e}+\vec p_T^{\, {\rm jet}}|\ll p_T^{\, e}\sim p_T^{\, {\rm jet}}$~\cite{Liu:2018trl}. For an incoming transversely polarized proton, the transverse spin vector $\vec s_T$ is correlated with the imbalance momentum $\vec q_T$, which leads to a $\sin(\phi_s - \phi_q)$ modulation of the electron-jet cross section~\cite{Liu:2018trl}. 
The spin-dependent differential cross section can be written as
\begin{align}
\frac{\diff \sigma(\vec s_T)}{\diff {\cal PS}} = F_{UU} + \sin(\phi_s - \phi_q) F_{UT}^{\sin(\phi_s - \phi_q)}\,,
\end{align}
where $\diff {\cal PS} = \diff y_e\,\diff^2 \vec p^{\; e}_T\,\diff^2 \vec q_T$, and $F_{UU}$ and $F_{UT}^{\sin(\phi_s - \phi_q)}$ are the unpolarized and transversely-polarized structure functions. The Sivers asymmetry is then given by
\begin{align}\label{eq:sivers-ratio}
A_{UT}^{\sin(\phi_s - \phi_q)} =\frac{F_{UT}^{\sin(\phi_s - \phi_q)}}{F_{UU}}\,.
\end{align}
Using TMD factorization in Fourier-transform space, the unpolarized differential cross section for electron-jet production can be written as
\begin{align}\label{eq:crosssection1}
    &F_{UU} = \sigma_0 \,H_q(Q,\mu)\sum_q e_q^2\, J_q(p_T^{\rm jet}R,\mu)
    \nn\\ & \times\,
    \int\frac{\diff^2 \vec b_T}{(2\pi)^2}\, e^{i\vec q_T\cdot \vec b_T}\, f_q^{\rm TMD}(x,\vec b_T,\mu)\, S_{q}(\vec b_T,y_{\rm jet},R,\mu) \,,
\end{align}
where $H_q$ is the hard function taking into account virtual corrections at the scale $Q$. The jet function $J_q$ takes into account the collinear dynamics of the jet formation which depends on the jet algorithm and the jet radius. Throughout this work we use the anti-$k_T$ algorithm~\cite{Cacciari:2008gp} and $R=1$. The quark TMD PDF  $f_q^{\rm TMD}$ includes the appropriate soft factor to make it equal to the one that appears in the SIDIS factorization~\cite{Collins:2011zzd}. The remaining soft function $S_{q}$ includes the contributions from the global soft function which depends on Wilson lines in the beam and jet direction, as well as the collinear-soft function that takes into account soft radiation along the jet direction. We summarize the fixed order results for the different functions in Appendix~\ref{app:one}. In addition, we include non-global logarithms~\cite{Dasgupta:2001sh} to achieve next-to-leading logarithmc (NLL$'$) accuracy. See also Refs.~\cite{Banfi:2003jj,Liu:2018trl,Buffing:2018ggv,Chien:2019gyf} for more details.

The Sivers structure function $F_{UT}^{\sin(\phi_s - \phi_q)}$ can be obtained from Eq.~(\ref{eq:crosssection1}) by replacing the usual unpolarized TMD PDF $f_q(x, k_T)$ with the quark Sivers distribution $f_{1T}^{\perp\,q}$ in the momentum space and by performing the corresponding Fourier transform to $b_T$-space
\begin{align}
    f_q(x, k_T) \to \frac{1}{M} \epsilon_{\alpha\beta} \,s_T^\alpha \,k_T^\beta \,f_{1T}^{\perp\, q}(x, k_T)\,,
\end{align}
where $M$ is the mass of the incoming proton. The remaining soft function $S_{q}$ is the same in the polarized and unpolarized case. Therefore, its nonperturbative contributions largely cancel in the Sivers asymmetry in Eq.~\eqref{eq:sivers-ratio}. In addition, the nonperturbative contribution of $S_{q}$ is expected to be subleading compared to the TMD PDF. Final-state hadronization effects are also expected to be small for the jet radius of $R=1$, which we choose for our numerical results presented below. 
The numerical size of the non-global logarithms is relatively small in the unpolarized case ${\cal O}(<4\%)$ and is expected to largely cancel out for the asymmetries we consider. 
Therefore, we expect that the cross section here provides a very clean handle on the quark TMD Sivers function.

\subsection{The Collins asymmetry and jet substructure}

Next, we consider the measurement of hadrons inside jets which is sensitive to the Collins TMD FF in the polarized case. In back-to-back electron-jet production, we now also include the hadron distribution inside the jet,
\begin{align}
    e+p(\vec s_T)\to e+({\rm jet}(\vec q_T)\,h(z_h, \vec j_T))+X\,. 
\end{align}
Here we consider both the longitudinal momentum fraction $z_h = \vec{p}_T^{\, h} \cdot \vec{p}_T^{\,\rm jet}/|\vec{p}_T^{\, \rm jet}|^{2}$ and the transverse momentum $\vec j_T = \vec{p}_T^{\, h} \times \vec{p}_T^{\, \rm jet}/|\vec{p}_T^{\, \rm jet}|^{2}$ of the hadron with respect to the jet axis. In this case, the spin-vector $\vec s_T$ of the incoming proton correlates with $\vec j_T$, which leads to a $\sin(\phi_s - \phi_h)$ modulation usually referred to as the Collins asymmetry for hadron in-jet production. Following the work of~\cite{Kaufmann:2015hma,Kang:2016ehg,Bain:2016rrv,Kang:2017glf,Ellis:2010rwa,Yuan:2007nd,Kang:2017btw,Kang:2019ahe,Kang:2020xyq}, the relevant cross section can be written as
\begin{align}\label{eq:structure}
    \frac{\diff \sigma^h(\vec s_T)}{\diff{\cal PS}\,\diff z_h\,\diff^2 \vec j_T^{\, h}} = 
    F_{UU}^h + \sin(\phi_s-\phi_h) F_{UT}^{\sin(\phi_s - \phi_h)}\,,
\end{align}
where $\phi_s$ is the azimuthal angle of the transverse spin of the incoming proton relative to the reaction plane and $\phi_h$ is the azimuthal angle of the hadron inside the jet. The Collins asymmetry for hadron in-jet is then given by
\begin{align}
    A_{UT}^{\sin(\phi_s - \phi_h)} = \frac{F_{UT}^{\sin(\phi_s - \phi_h)}}{F_{UU}^h}\,. 
\end{align}
The unpolarized structure function $F_{UU}^h$ for hadron in-jet production is given by
\begin{align}\label{eq:crosssection2}
    &F_{UU}^h =
    \sigma_0 \,H_q(Q,\mu) \sum_q e_q^2\, {\cal G}_q^h(z_h,\vec j_T,p_T^{\rm jet}R,\mu)
    \nn\\ &\times \,
    \int\frac{\diff^2 \vec b_T}{(2\pi)^2}\, e^{i\vec q_T\cdot \vec b_T}\, f_q^{\rm TMD}(x,\vec b_T,\mu)\, S_{q}(\vec b_T,y_{\rm jet},R,\mu) \,.
\end{align}
Here ${\cal G}_q^h$ is a TMD fragmenting jet function~\cite{Kang:2017glf,Kang:2019ahe,Procura:2009vm} which captures the dependence on the jet substructure measurement. It replaces the jet function $J_q$ in Eq.~(\ref{eq:crosssection1}) and satisfies the same renormalization group evolution equation. At the jet scale $p_T^{\rm jet}R$, up to NLL we can write ${\cal G}_q^h$ in Fourier space as~\cite{Kang:2019ahe,Kang:2017glf}
\begin{equation}\label{eq:TMDFF}
    {\cal G}_q^h(z_h,\vec j_T,p_T^{\rm jet}R)=\int\frac{\diff^2\vec b_T^{\,\prime}}{(2\pi)^2}e^{i\vec j_T\cdot\vec b_T^{\,\prime}/z_h} D_{h/q}^{\rm TMD}(z_h,\vec b_T^{\,\prime},p_T^{\rm jet}R) \,.
\end{equation}
Here $D_{h/q}^{\rm TMD}$ is the unpolarized TMD FF evaluted at the jet scale. The superscript ``TMD'' indicates that we have included the proper soft function to make it equal to the standard TMD FFs as probed in SIDIS and/or in back-to-back dihadron production in $e^+e^-$ annihilation. We use the Fourier variable $\vec b_T^{\,\prime}$ to indicate that this integration is independent of the TMD PDF in Eq.~(\ref{eq:crosssection2}).

The spin-dependent structure function $F_{UT}^{\sin(\phi_s - \phi_h)}$ is obtained from Eq.~(\ref{eq:crosssection2}) by replacing the unpolarized TMD PDF $f_q$ with the TMD quark transversity distribution $h_1^q$, the unpolarized TMD FF $D_{h/q}$ with the Collins TMD fragmentation function $H_{1}^{\perp\,q}$, and using the appropriate polarized cross section $\sigma_0^{\rm Collins}$. We thus have
\begin{align}
    f_q(x,k_T) &\to h_1^q(x, k_T)\,,
    \\
    D_{h/q}(z_h, j_T) &\to \frac{j_T}{z_h M_h} H_{1 h/q}^{\perp}(z_h, j_T)\,,
    \\
    \sigma_0 &\to \sigma_0^{\rm Collins} = \frac{\alpha\alphas}{s Q^2}\frac{4\hat s \hat u}{-\hat t^2}\,,
\end{align}
where $M_h$ is the mass of the observed hadron in the jet. See Ref.~\cite{Kang:2017btw} for more details.

\section{Simulation~\label{sec:simulation}}

We use simulations to explore the kinematic reach and statistical precision subject to the expected acceptance of EIC experiments, as well as to estimate the impact of the detector resolution. We use \textsc{Pythia8}~\cite{Sjostrand:2007gs} to generate neutral-current DIS events in unpolarized electron-proton collisions, see Fig.~\ref{fig:NC_DIS}. \textsc{Pythia8} uses leading-order matrix elements matched to the \textsc{DIRE} dipole shower~\cite{Hoche:2015sya}, and subsequent Lund string hadronization. For consistency with the calculations presented in Section~\ref{sec:theory}, we do not include QED radiative corrections in the simulation.

We set the energies of the electron and proton to 10~GeV and 275~GeV, respectively. These beam-energy values, which yield a center-of-mass energy of {$\sqrt{s}=105$~GeV}, correspond to the operation point that maximizes the luminosity in the eRHIC design~\cite{EICdesign}.  We consider yields that correspond to an integrated luminosity of 100~fb$^{-1}$, which can be collected in about a year of running at 10$^{34}$~cm$^{-2}$s$^{-1}$.

We select events with $Q^{2} > 25$ \GeVsquare~and $0.1<y<0.85$. The lower elasticity limit avoids the region where the experimental resolution of the DIS kinematic variables $x$ and $Q^{2}$ diverges and the upper limit avoids the phase space in which QED radiative corrections are significant.

We do not simulate jet photo-production, which is a negligible contribution at high $Q^{2}$~\cite{Abelof:2016pby,Frank}. By lowering $Q^2$ and including photo-production, the jet rate would increase, but at the cost of sensitivity to photon PDFs~\cite{Matevosyan:2015gwa}. See, for example, Refs.~\cite{Jager:2008qm,Aschenauer:2019uex} for EIC studies of jets in photo-production events and Refs.~\cite{Kang:2011jw,Hinderer:2015hra,Abelof:2016pby} where the entire $Q^2$ range is considered.

We use the \textsc{Fastjet}3.3 package~\cite{Cacciari:2011ma} to cluster jets with the anti-$k_T$ algorithm and radius parameter $R=1.0$. HERA studies showed that such a large value of $R$ reduced hadronization corrections for inclusive jet spectra to the percent level~\cite{Newman:2013ada}. The input for the jet clustering algorithm are stable particles that have transverse momentum $p_T>100$~MeV and pseudorapidity $|\eta|<4.0$ in the laboratory frame\footnote{Throughout this paper, we follow the HERA convention to define the coordinate system. The $z$ direction is defined along the proton beam axis and the electron beam goes toward negative $z$. The polar angle $\theta$ is defined with respect to the proton (ion) direction.}, excluding neutrinos and the scattered electron\footnote{We identify the scattered electron as the electron with the largest $p_T$~in the event.}. 

Unlike most projection studies for the EIC, we do not use the Breit frame but instead use the laboratory frame. This approach was advocated for by Liu et al.~\cite{Liu:2018trl} in order to have a close connection to results from hadron colliders, such as di-jet studies~\cite{Abelev:2007ii,Boer:2009nc}.  As discussed in Ref.~\cite{Arratia:2019vju}, this is not a trivial change of reference frame because a low $p_T$~threshold would suppress most of leading-order DIS events (called ``quark-parton-model background'' in most HERA jet studies~\cite{Newman:2013ada}).

We impose a minimum cutoff of 5~GeV in transverse momentum for both the electron and jet to ensure a reasonable prospect of reconstruction efficiency as well as to provide a scale to control perturbative QCD calculations. 

\begin{figure}
    \centering
    \includegraphics[width=0.49\textwidth]{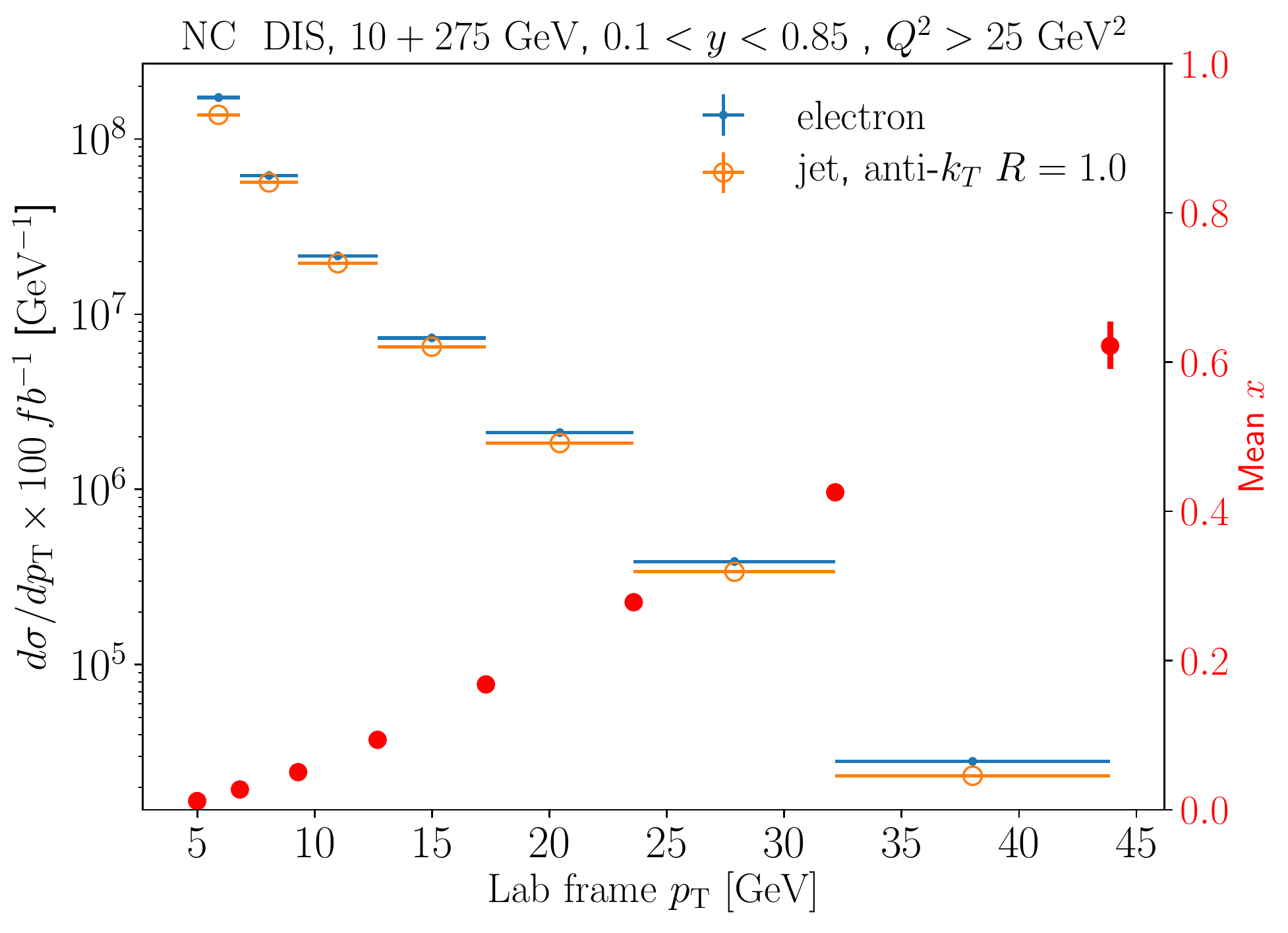}
\caption{Yield of electrons and jets, and mean $x$ as a function of the transverse momentum in the laboratory frame. The jets were reconstructed with the anti-$k_T$~algorithm~\cite{Cacciari:2008gp} and $R=1$. The red error bars represent the standard deviation of the $x$ distribution for each electron $p_T$~interval.~\label{fig:xsection}}
\end{figure}

Figure~\ref{fig:xsection} shows the differential yield of electrons and jets and the probed average $x$ value as a function of $p_T$ in the lab frame. The yield of electrons and jets are similar at high \pT, as expected from leading-order DIS, whereas they differ at low $p_T$~due to parton branching processes or out-of-jet emission, and hadronization effects. We have verified that the \textsc{Pythia8} cross section is within 5$\%$ of next-to-next-to-leading order pQCD calculations~\cite{Abelof:2016pby,Frank}, which is sufficient for our estimates.  

The sea-quark-dominated region is probed with low-$p_T$~jets, $x\approx 0.05$ at $p_T\approx 7$~GeV. The valence region, $x>0.1$, is reached with $p_T\sim 15$ GeV and the region $x>0.3$, which remains unconstrained for transversely-polarized collisions~\cite{Anselmino:2020vlp}, is probed with $p_T>25$ GeV.

While 100 fb$^{-1}$ of integrated luminosity would provide more than enough statistics for precise cross-section measurements over the entire $p_T$~range, the high luminosity will be critical for for multi-dimensional measurements and to constrain the small transverse-spin asymmetries expected for EIC kinematics, as we show in the next section. 

\section{Numerical results and statistical projections~\label{sec:projections}}

In this section, we present numerical results using the theoretical framework presented in  Section~\ref{sec:theory} and we estimate the statistical precision of future measurements at the EIC. 

\subsection{Unpolarized production of jets \\ and jet substructure}

Before presenting the results for the asymmetry measurements, we first compare our numerical results for jets and jet substructure in unpolarized electron-proton collisions to \textsc{Pythia8} simulations. 

\begin{figure}
    \centering
    \includegraphics[width=0.49\textwidth]{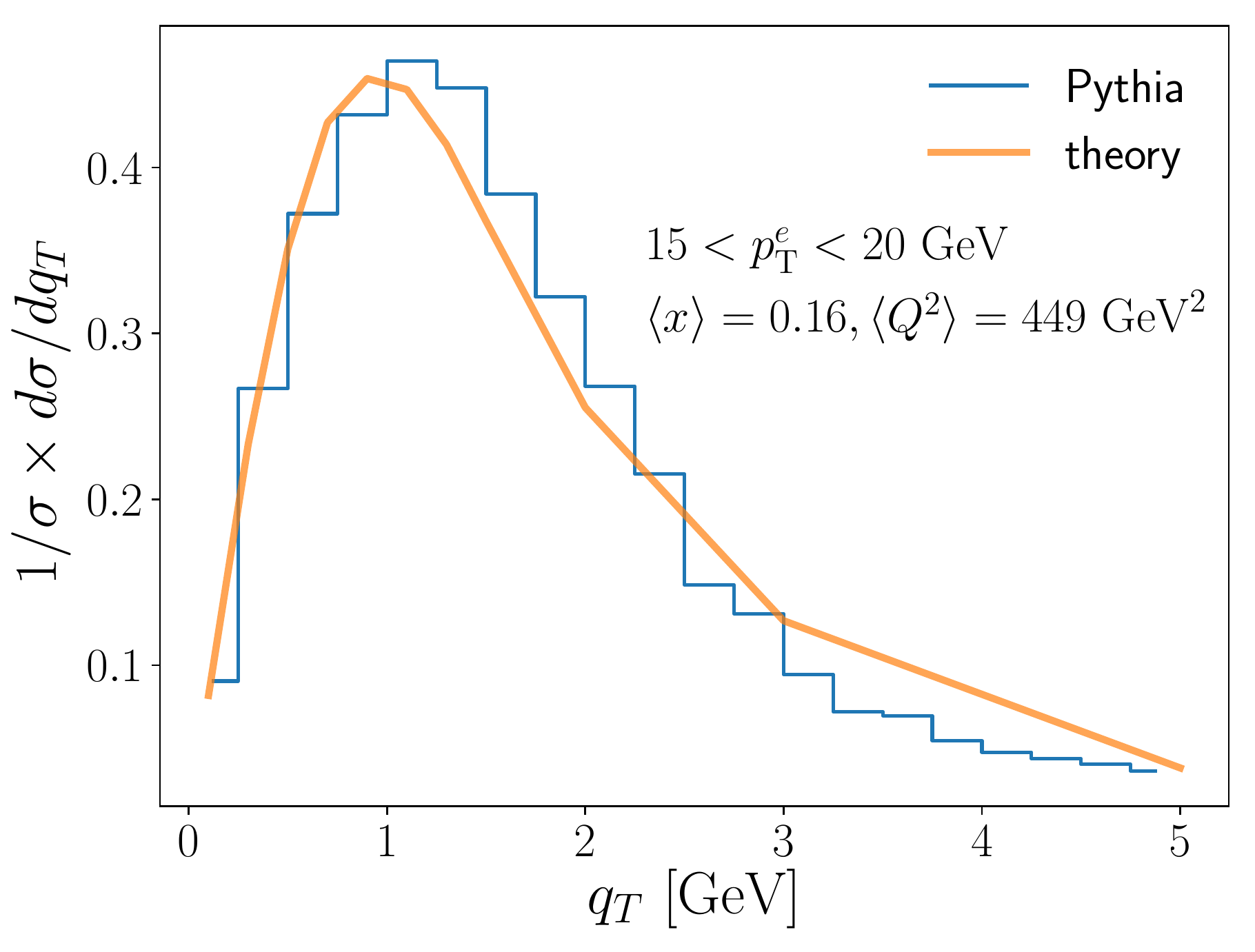}
\caption{Normalized distribution of the transverse momentum imbalance $q_T$ for jets produced in unpolarized electron-proton collisions. We integrate over the event inelasticity $0.1 < y < 0.85$ and electron transverse momentum $15<p_T^{e}<20$ GeV.} \label{fig:qt}
\end{figure}

We start with the electron-jet production. Figure~\ref{fig:qt} shows the normalized distribution of the transverse momentum $q_T$ for jets produced in unpolarized electron-proton collisions. We integrate over the event inelasticity $0.1 < y < 0.85$ and electron transverse momentum $15<p_T^{e}<20$ GeV. The distribution shows the expected Gaussian-like behavior at small values of $q_T\lesssim 2$ GeV, which is driven by the TMD PDF and soft gluon radiation, and a tail to intermediate values of $q_T$, which is driven by perturbative QCD radiation. We observe a reasonable agreement with the \textsc{Pythia8} results.

\begin{figure*}
    \centering
   \includegraphics[width=0.49\textwidth]{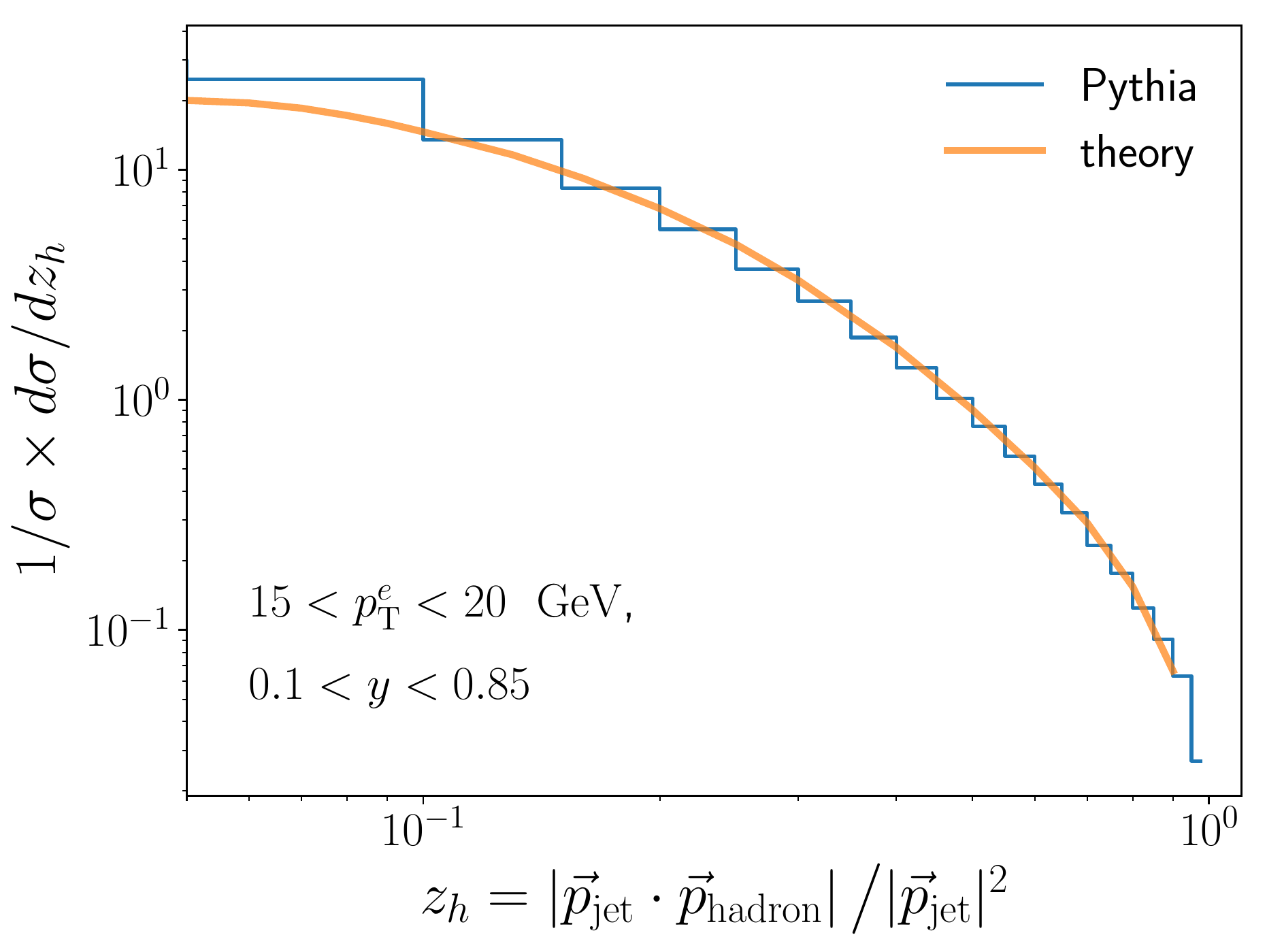}
    \includegraphics[width=0.49\textwidth]{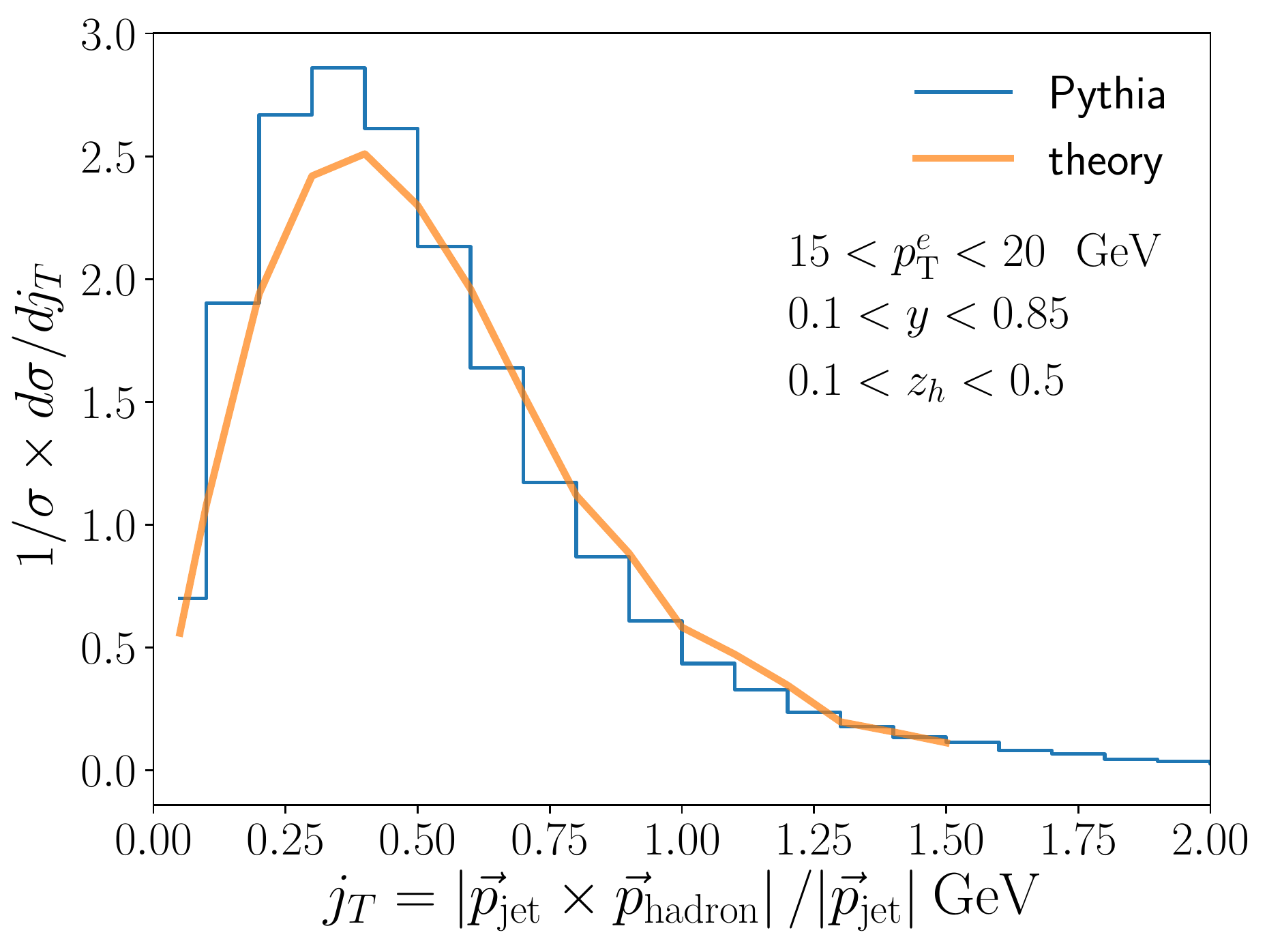}
\caption{Numerical results using our theoretical framework (orange) and \textsc{Pythia8} calculations (blue histograms) for the longitudinal momentum fraction $z_h$ (left panel) and the transverse momentum $j_T$ (right panel) for charged hadrons inside jets at the EIC. The results shown here are for the unpolarized case. We also include a cut of $q_T/p_T^{\rm jet}<0.3$ as discussed in the text.~\label{fig:zj}}
\end{figure*}

We now turn to the jet substructure results, for which we impose a selection cut of $q_T/p_T^{\rm jet} < 0.3$ to ensure the applicability of the TMD framework. Fig.~\ref{fig:zj} shows the hadron-in-jet distributions as a function of $z_h$ integrated over $j_T$, as well as the $j_T$ distribution integrated over $0.1 < z_h < 0.5$. We use the DSS fit of the collinear FFs of Ref.~\cite{deFlorian:2007aj}, while the TMD parametrization is taken from Ref.~\cite{Kang:2015msa}. We observe a very good agreement for the $z_h$ distribution and the \textsc{Pythia8} simulation, and a reasonable agreement for the $j_T$ distribution. In the absence of experimental data, these results provide confidence in our theoretical framework. 

\subsection{Spin asymmetries}

Here, we study spin asymmetries in the collisions of electrons and transversely-polarized protons. Given that most of the systematic uncertainties cancel in the asymmetry measurements, statistical uncertainties will likely dominate the total uncertainties. We estimate the impact of detector resolution and other requirements in Section~\ref{sec:detector}. 

We estimate the statistical uncertainties of the asymmetry measurements assuming an integrated luminosity of 100 fb$^{-1}$ and an average proton-beam polarization of $70\%$, following the EIC specifications~\cite{Accardi:2012qut}. We also assume a conservative value of 50$\%$ for the overall efficiency due to the trigger efficiency, data quality selection, and reconstruction of electrons, and jets. For small values of the asymmetry, the absolute statistical uncertainty can be approximated as $\delta A \approx 1/(\sqrt{N} p$), with $p$ is the average nucleon polarization and $N$ the yield summed over polarization states. For the Collins asymmetry, we also include a penalty factor of $\sqrt{2}$, which arises from the statistical extraction of simultaneous modulations of the hadron azimuthal distribution~\cite{Anselmino:2011ay}. We also estimate the increase of statistical uncertainty due to ``dilution factors'' caused by smearing in either the Sivers angle (azimuthal direction of $\vec{q_{T}}$) or the Collins angle (azimuthal direction of $\vec{j_{T}}$); these are described in Section~\ref{sec:detector}. 

\subsubsection{Electron-jet azimuthal correlations~\label{sec:Sivers}}

\begin{figure}
    \centering
    \includegraphics[width=0.49\textwidth]{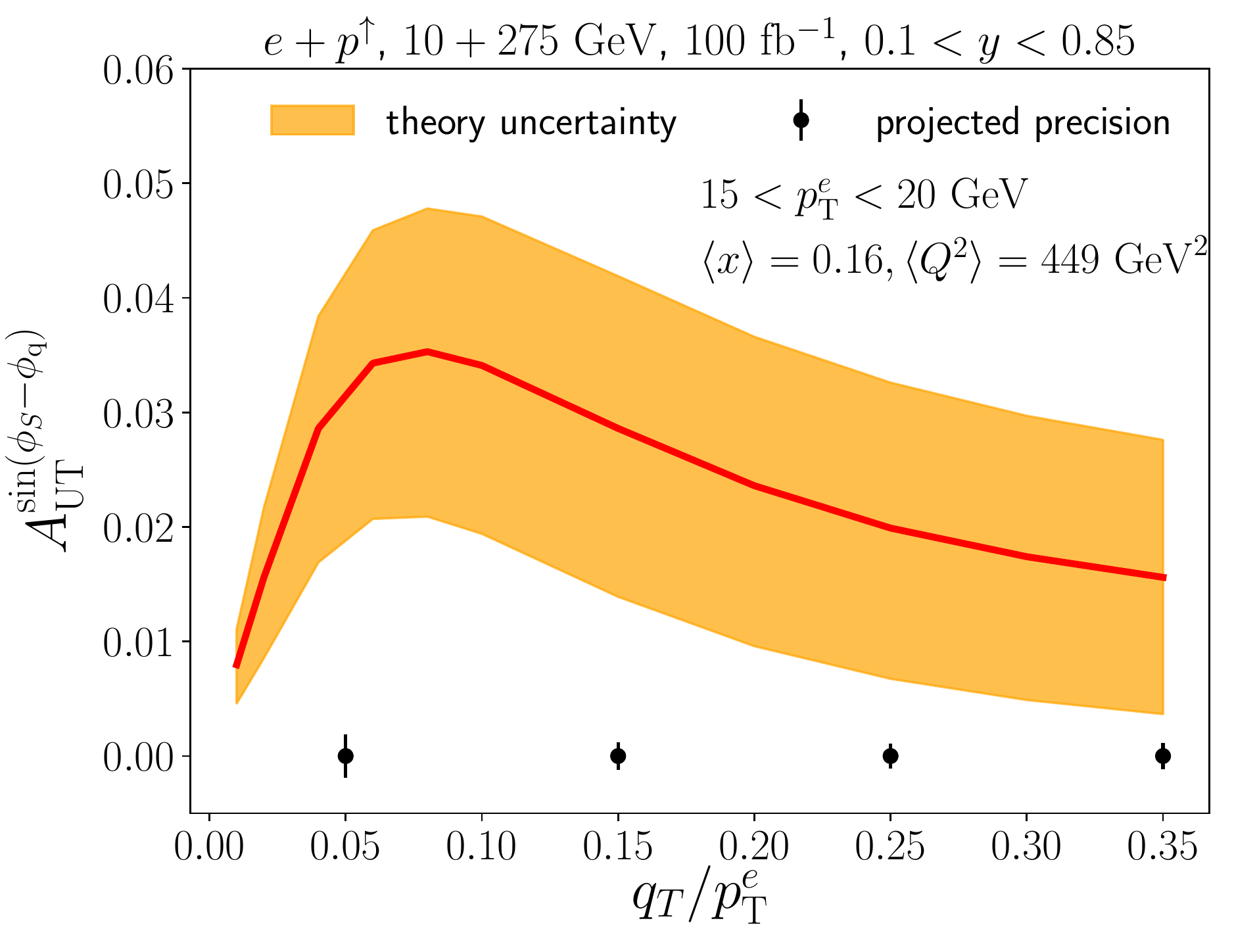}
\caption{Theoretical result for the electron-jet asymmetry sensitive to the Sivers distribution (red). The uncertainty band (orange) displays the current uncertainty of the Sivers function of Ref.~\cite{Echevarria:2014xaa}. In addition, we show projections of statistical uncertainties for an EIC measurement (black error bars).~\label{fig:Sivers}}
\end{figure}

We start with the Sivers asymmetry which is accessed through the measurement of the electron-jet correlation. Fig.~\ref{fig:Sivers} shows numerical results for the Sivers asymmetry $A_{UT}^{\sin(\phi_s-\phi_q)}$ in Eq.~\eqref{eq:sivers-ratio} including an uncertainty band according to the extraction of Ref.~\cite{Echevarria:2014xaa}. In addition, we show the projected statistical uncertainty of the Sivers asymmetry measurement as a function of $q_T/p_T^{\rm e}$. We integrate again over $15 < p_T^e < 20$ GeV and $0.1 < y < 0.85$, and thus the probed $x$ range for the quark Sivers function is integrated over. The theoretical uncertainty is calculated solely based on the uncertainty of the extracted quark Sivers function~\cite{Echevarria:2014xaa} from current SIDIS measurements; other extractions of the Sivers function~\cite{Bacchetta:2020gko} are expected to lead to similar uncertainty. The projected statistical uncertainty is much smaller than the theoretical uncertainty, which implies that the EIC jet measurements will
help to better constrain the quark Sivers function.

While most systematic uncertainties cancel in the ratio of the asymmetry, including jet-energy scale and jet-energy resolution uncertainties, the differential measurement of the Sivers asymmetry demands resolution on the $q_T/p_T^{e}$ measurement. We address this issue in Section~\ref{sec:detector}. 

The hard scale at which the jet-based Sivers measurement can be performed is much closer to analogous Drell-Yan measurements at RHIC~\cite{Aschenauer:2016our}. This 
would lead to a better handle on TMD evolution effects, which ultimately can help confirm the sign-change of the Sivers function between SIDIS and Drell-Yan reactions~\cite{Brodsky:2002cx,Collins:2002kn,Boer:2003cm}.

\subsubsection{Hadron-in-jet asymmetries~\label{sec:hadroninjet}}

\begin{figure*}
    \centering
    \includegraphics[width=0.95\textwidth]{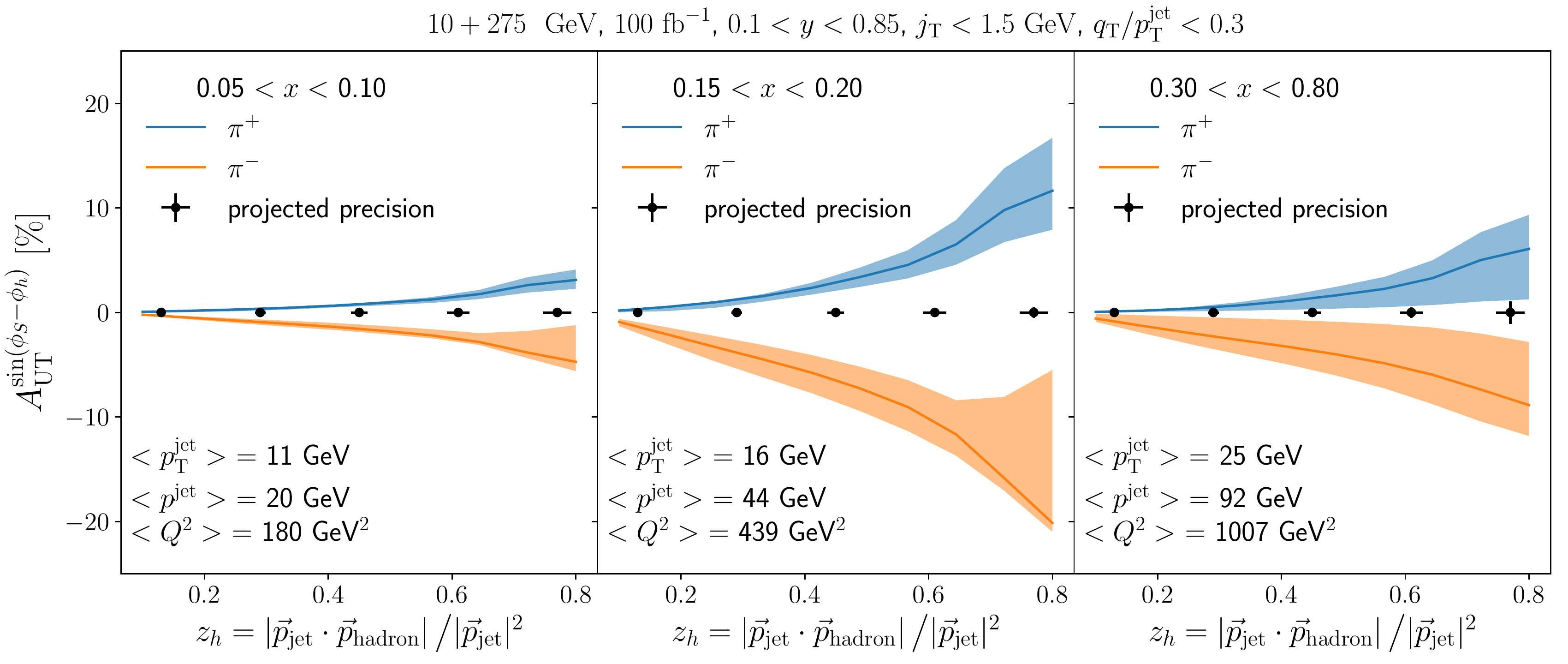}
\caption{Projection of statistical uncertainties (black error bars) for the $z_h$ distribution for the $\pi^{\pm}$-in-jet Collins asymmetries as well as theoretical predictions (blue, orange). The displayed theoretical uncertainties (orange and blue bands) are based on the extraction of Ref.~\cite{Kang:2015msa}. The horizontal error bar corresponds to a jet-energy scale uncertainty of 3$\%$. }
\label{fig:Collins}
\end{figure*}

Next, we are going to study the Collins asymmetry via the distribution of hadrons inside the jet. Figure~\ref{fig:Collins} shows the projected precision for three $x$ intervals: $0.05 < x < 0.1$, $0.15 < x < 0.2$, and $0.30 < x < 0.80$, along with our theoretical calculations for the in-jet Collins asymmetry for $\pi^+$ and $\pi^-$ as a function of $z_h$. The projected precision assumes a fully-efficient identification for $\pi^{\pm}$ with negligible misidentification with other hadron species; we discuss the requirements for particle-identification systems in Section~\ref{sec:detector}. The theory uncertainty bands are obtained from the quark transversity and Collins fragmentation functions extracted in Ref.~\cite{Kang:2015msa}. The extraction from  Ref.~\cite{Kang:2015msa} is based on a simultaneous fit of the SIDIS Collins asymmetry and the Collins asymmetry in back-to-back hadron pair production in $e^+e^-$ collisions. The projected statistical uncertainties at the EIC are much smaller than the uncertainties obtained from current extractions. Therefore, future in-jet Collins asymmetry measurements at the EIC will provide important constraints on both the quark transversity and the Collins fragmentation functions. 

The region $x < 0.1$ (relevant for sea quarks) is not well known from current SIDIS measurements. The measurements at the EIC will provide excellent constraining power for the sea-quark distribution. The projected uncertainties in the valence-dominated region are larger, but still provide enough sensitivity compared to the predicted asymmetries. These measurements will complement future measurements from SoLID~\cite{Chen:2014psa} and the STAR~\cite{Aschenauer:2015eha} experiment. 

Impact studies of the projected EIC data on quark transversity, similar to Ref.~\cite{Ye:2016prn}, are beyond the scope of this work but will be addressed in future publications.

\section{Detector performance ~\label{sec:detector}}

In this section, we estimate the detector performance for electron-jet and hadron-in-jet measurements. The measurement of the scattered electron defines the inclusive DIS measurement and has been discussed in  detail~\cite{EICHandbook}, so we focus on jets.

\subsection{Jet kinematics}

\begin{figure*}[t]
    \centering
   \includegraphics[width=0.33\textwidth]{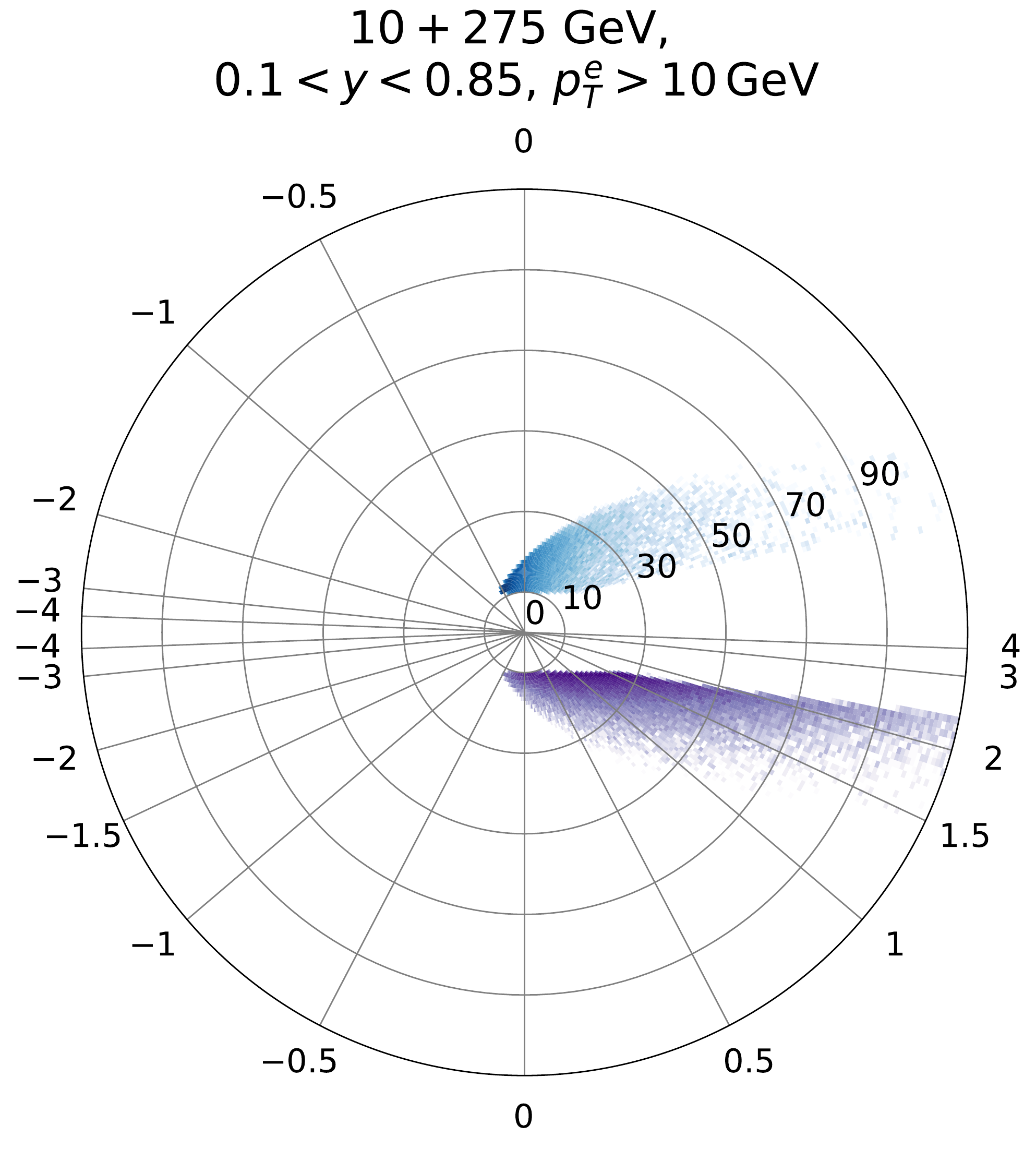}
    \includegraphics[width=0.33\textwidth]{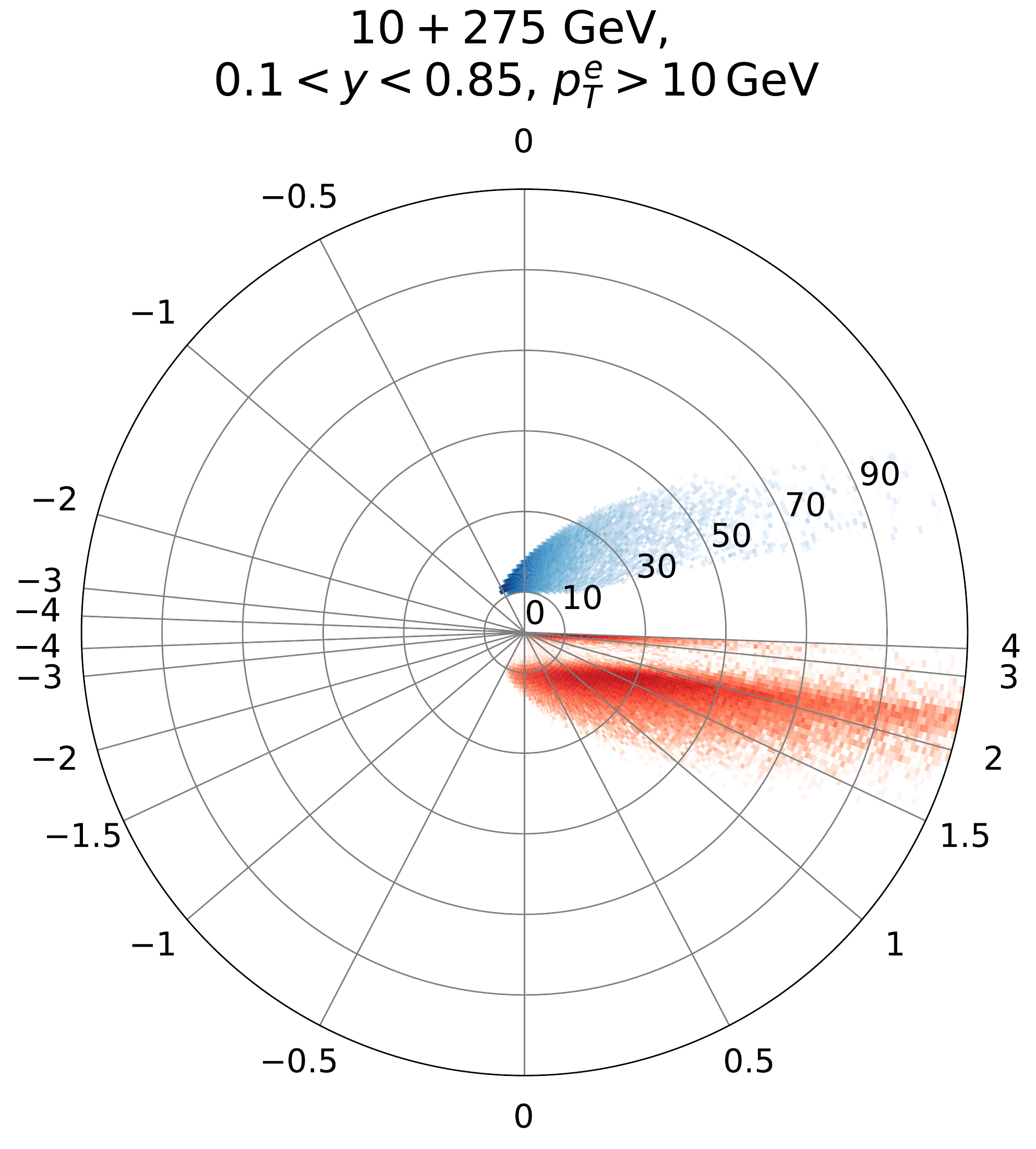}
    \caption{The top half of each circle shows the pseudorapidity and 3-momentum of the scattered electron in the angular and radial direction, respectively. The bottom half of each circle shows the pseudorapidity and momentum of the struck quark (left) and jets (right). The jets were reconstructed with the anti-$k_{\mathrm{T}}$ algorithm and $R=1.0$.}
    \label{fig:polarplot}
\end{figure*}

Figure~\ref{fig:polarplot} shows the momentum and pseudorapidity distribution of electrons (upper half plane), the struck quark and jets (lower half plane). The jet distribution matches the struck-quark kinematics to a remarkable degree.  The polar plot on the right includes initial and final-state radiation, hadronization, and the beam remnants.

For this very asymmetric beam-energy configuration (10 GeV electron and 275 GeV proton) jets are predominantly produced around $\eta\approx 1.5$. The larger the $x$ of the event, the more forward is the jet. While some of the jets are at mid-rapidity $(-1.0<\eta<1.0)$, they are predominantly produced in the challenging region between the barrel and endcap of a typical collider detector. Given that large-$R$ jets are preferred to minimize hadronization corrections associated with the jet clustering algorithm, this will impose a challenge for the detector design. While acceptance gaps and dead material due to services are inevitable, they should be limited to not compromise the acceptance of large-$x$ events, which is where the Sivers and transversity functions have maximums. Gaps in acceptance, particularly in calorimeters, would lead to a mismeasurement of the jet energy that would require corrections sensitive to modeling of hadronization (event generator) and detector effects (detailed geometry and material description). 

\subsection{Fast simulations}

We use the \textsc{Delphes} package of Ref.~\cite{deFavereau:2013fsa} for fast detector simulations. We consider the geometry of a general-purpose collider detector: tracking, electromagnetic and hadronic calorimeters with hermetic coverage in pseudorapidity up to $|\eta|=4$ and full azimuthal coverage. The parametrization of momentum and energy resolutions used as input for \textsc{Delphes} are shown in Table~\ref{tab:resolutions}. These values closely follow the requirements for a general-purpose detector at the EIC~\cite{EICHandbook}, and are the same as used in Ref.~\cite{Arratia:2020azl}. While these parameters are preliminary and subject to change given ongoing studies, they are a reasonable choice for our feasibility studies. 

\begin{table}[h!]
\centering
\caption{Parametrization of the momentum and energy resolution used as input for the \textsc{Delphes} fast simulations. These follow closely the baseline for an EIC general-purpose detector in Ref.~\cite{EICHandbook}.}
 \begin{tabular}{l l} 
 \hline
        Tracker, ${\rm d}p/p$ &0.5\% $\oplus$ 0.05\%$\times p$ for $|\eta|<1.0$ \\       
                        &  1.0\% $\oplus$ 0.05\%$\times p$ for $1.0<|\eta|<2.5$ \\ 
                        & 2.0\% $\oplus$ 0.01\%$\times p$ for $2.5<|\eta|<3.5$ \\
                        \hline 
        EMCAL, ${\rm d}E/E$   &  2.0\%/$\sqrt{E}$ $\oplus$ 1\% for $-$3.5$<\eta<$2.0 \\
                        &   7.0\%/$\sqrt{E}$ $\oplus$ 1\% for $-$2.0$<\eta<-$1.0 \\
                        &   10.0\%/$\sqrt{E}$ $\oplus$ 2\% for $-$1.0$<\eta<-$1.0 \\
                         &   12\%/$\sqrt{E}$ $\oplus$ 2\% for 1.0$<\eta<$3.5 \\
        \hline
        HCAL, ${\rm d}E/E$   &  100\%/$\sqrt{E}$ $\oplus$ 10$\%$ for $|\eta|<$1.0 \\
               &   50\%/$\sqrt{E}$ $\oplus$ 10$\%$ for 1.0$<|\eta|<$4.0 \\
               \hline
\end{tabular}
\label{tab:resolutions}
\end{table}

\textsc{Delphes} implements a simplified version of the particle-flow algorithm to reconstruct jets, missing-energy, electrons, and other high-level objects. This algorithm combines the measurements from all subdetectors. While the fast simulation in \textsc{Delphes} lacks a detailed description of hadronic and electromagnetic showers, it approximates well the jet and missing-transverse-energy performance obtained with a \textsc{Geant}-based simulation of the CMS detector~\cite{Agostinelli:2002hh}, even down to 20 GeV. 

Table~\ref{tab:granularity} shows the granularity used in the \textsc{Delphes} simulation. At mid rapidity, the granularity follows that of the sPHENIX hadronic calorimeter~\cite{Aidala:2017rvg}, which is currently under construction. In the forward-rapidity region ($1.0<|\eta|<4.0$), we consider a granularity that roughly corresponds to $10\times10$ cm$^{2}$ towers positioned at 3.5 m; the tower size follows the STAR forward-calorimeter technology~\cite{Tsai_2015}. No longitudinal segmentation is considered for the calorimeters, as it is currently beyond the scope of \textsc{Delphes}. 

\begin{table}[h!]
\centering
\caption{Calorimeter granularity parameters ($\Delta\eta \times \Delta\phi$, in radians) used as input for the \textsc{Delphes} fast simulations.}
 \begin{tabular}{l l} 
 \hline
        EMCAL  &  $0.020 \times 0.020$ for $|\eta|<$1.0 \\
                        &   $0.020 \times 0.020$ for for 1.0$<|\eta|<$4.0 \\
        HCAL  & $0.100 \times 0.100$ for $|\eta|<$1.0 \\
               &   $0.025 \times 0.025$ for $1.0<|\eta|<4.0$ \\
               \hline
\end{tabular}
\label{tab:granularity}
\end{table}

The calorimetric energy thresholds are set to 200~MeV for the EMCAL and 500~MeV for the HCAL, which is possible for the expected noise levels~\cite{Tsai_2015}. A minimum significance, $E/\sigma(E)>1.0$, is required. A minimum track $p_T$~of 200 MeV is considered. The tracking efficiency is assumed to be 100$\%$ with negligible fake rate. 

\textsc{Delphes} simulates the bending of charged particles in a solenoidal field, which is set to 1.5 T. The volume of the magnetic field is assumed to cover a radius of 1.4~m and a half-length of 1~m, which roughly follows the dimensions of the BaBar solenoid magnet that is currently being used for the sPHENIX detector~\cite{Adare:2015kwa}.   

\begin{figure*}
    \centering
   \includegraphics[width=1.0\columnwidth]{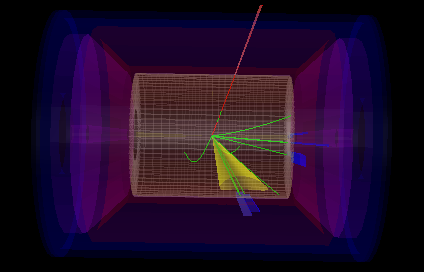}
    \includegraphics[width=0.75\columnwidth]{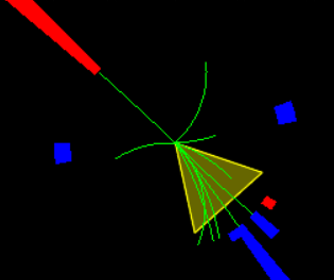}
   \caption{Left: event display showing a general-purpose EIC detector implemented in \textsc{Delphes} and a neutral-current DIS event at 105~GeV center-of-mass energy. Charged-particles are shown in green, hits in the electromagnetic calorimeter in red, hits in the hadronic calorimeter in blue. Right panel: event display with electron and jet in the back-to-back configuration studied in this work.~\label{fig:detector}}
\end{figure*}

Jets are reconstructed using \textsc{Delphes} particle-flow objects as inputs to the anti-$k_{T}$ algorithm~\cite{Cacciari:2008gp} with $R=1.0$ implemented in \textsc{Fastjet}~\cite{Cacciari:2011ma}. Given the relatively low energy of jets at the EIC and the superior tracking momentum resolution over the HCAL energy resolution, jets reconstructed with purely calorimetry information yield worse performance and are not considered here. 

Figure~\ref{fig:detector} shows an event display for a neutral-current DIS event reconstructed with the detector geometry described above. The signal for our studies is an isolated electron and a jet back-to-back in the transverse plane. The displayed event is representative for the particle multiplicity expected in high-$Q^{2}$ DIS events at the EIC~\cite{Page:2019gbf,Arratia:2019vju}. Very clean jet measurements will be possible given that underlying event and pileup will be negligible. As shown in Ref.~\cite{Arratia:2019vju}, the average number of particles in jets ranges from about 5 at $\pTjet=5$~GeV to about 12 at $\pTjet=25$~GeV. 

The jet performance is estimated by comparing jets ``at the generator level'' and at the ``reconstructed level''. The input for the jet clustering at the generator level are final-state particles in \textsc{Pythia8}, whereas the input for the reconstructed level are particle-flow objects from \textsc{Delphes}. Reconstructed jets are matched to the generated jets with an angular-distance selection of $\Delta R = \sqrt{(\phi_{\rm jet}^{\rm gen}- \phi_{\rm jet}^{\rm reco})^{2} + (\eta_{\rm jet}^{\rm gen}- \eta_{\rm jet}^{\rm reco})^{2}}  <0.3$, which is fully efficient for jets with $\pTjet>10$ GeV. 

\begin{figure}
    \centering
   \includegraphics[width=0.49\textwidth]{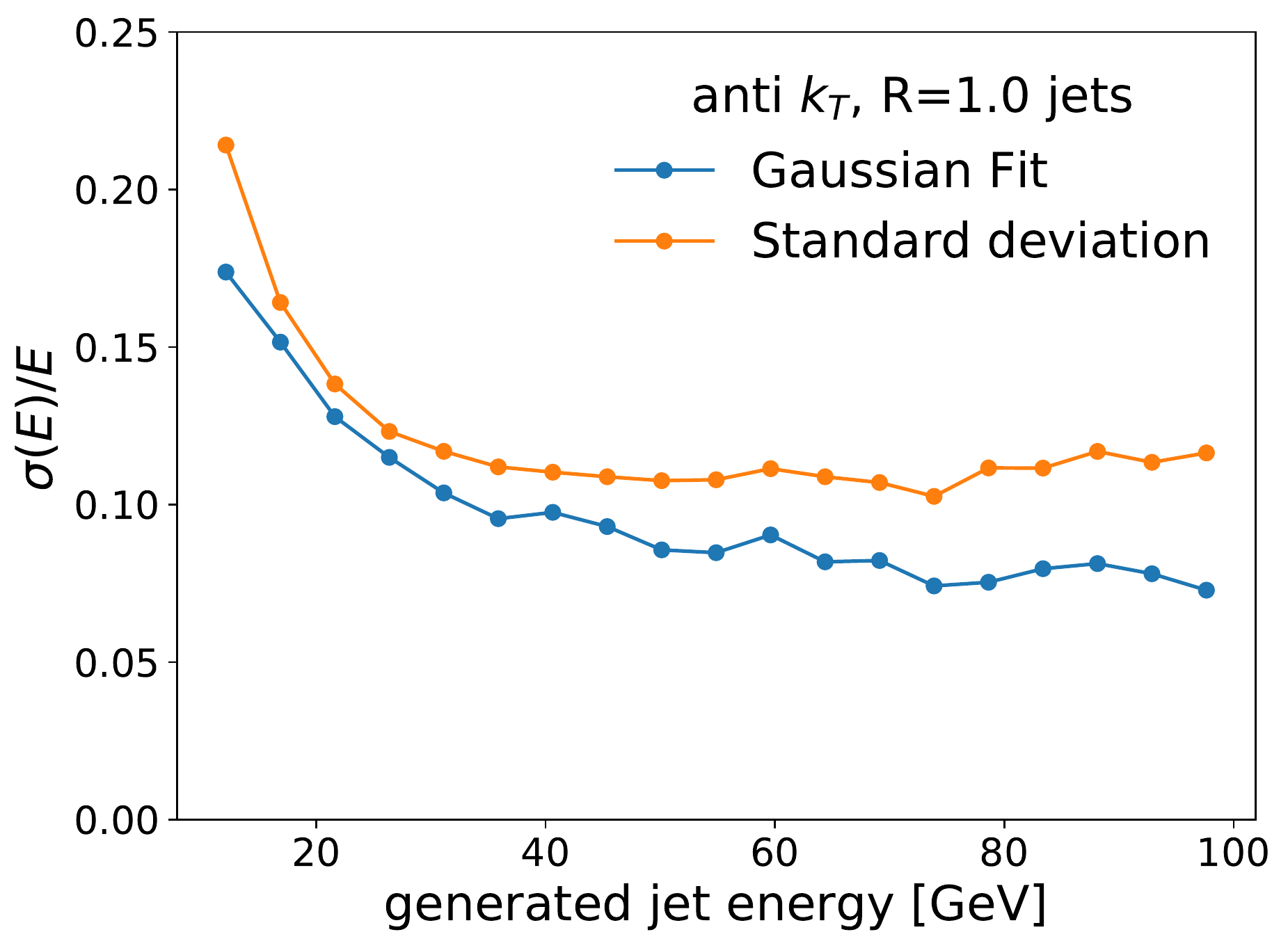}
   \caption{Relative energy resolution for jets produced in neutral-current DIS events. The jets are reconstructed with the anti-$k_{T}$ algorithm with $R=$1.0 using \textsc{Delphes} particle-flow objects.}
   \label{fig:jetperformance}
\end{figure}

Figure~\ref{fig:jetperformance} shows the jet resolution, which is defined by a Gaussian fit to the relative difference between generated and reconstructed jet momentum. The resolution is driven by the response of the calorimeters. The non-Gaussian tails of the detector response are quantified by comparing the jet-energy resolution estimated by computing a standard deviation instead of the Gaussian fits. The difference is about 1--4$\%$, which indicates that the response matrix does not have large non-diagonal elements, which appear in detector designs that do not consider a hadronic calorimeter in the barrel region, as noted by Page et al.~\cite{Page:2019gbf}. A diagonal response matrix (i.e. a Gaussian-like resolution) will enable accurate jet and missing-transverse energy measurements, see also Ref.~\cite{Arratia:2020azl}.  

\begin{figure}
    \centering
   \includegraphics[width=0.49\textwidth]{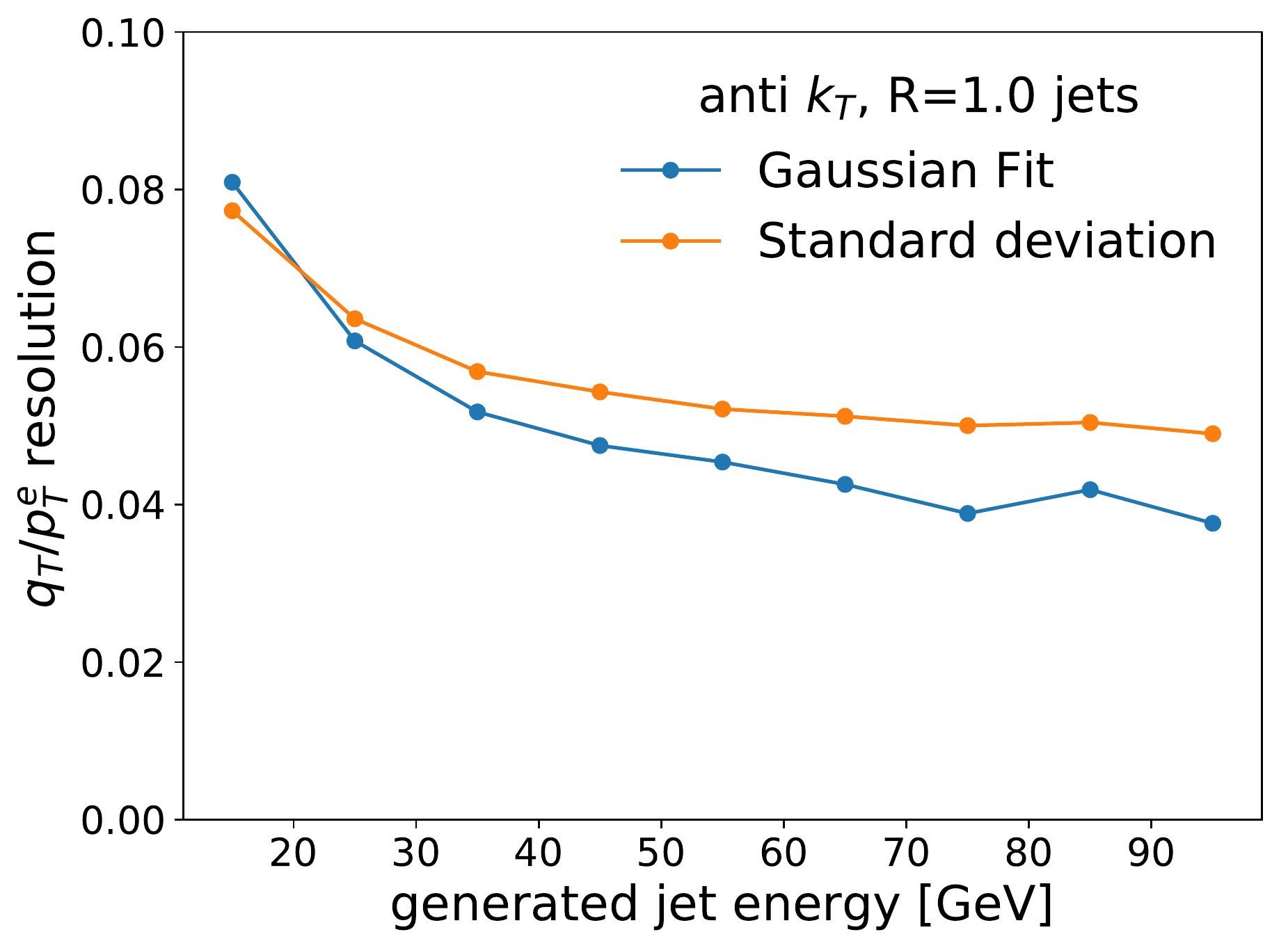}
   \caption{Absolute resolution for the normalized electron-jet imbalance, $q_{T}/p_{T}^{e}$, as a function of the generated jet energy. The jets are reconstructed with the anti-$k_{T}$ algorithm with $R=$1.0 using \textsc{Delphes} particle-flow objects.}
   \label{fig:azimuthalresponse}
\end{figure}

Figure~\ref{fig:azimuthalresponse} shows the expected resolution on the electron-jet azimuthal imbalance $q_{T}$ normalized by $p_{T}^{e}$. This resolution informs the bin-widths presented in Figure~\ref{fig:Sivers} to ensure controllable bin-migration; we leave detailed unfolding studies for future work. 

A better resolution could be achieved by defining $q_{T}$ with charged particles only, which would require us to introduce track-jet functions~\cite{Chang:2013iba, Chang:2013rca} in the theoretical framework as done in Ref.~\cite{Chien:2020hzh}.   

We find that the resolution of the Sivers angle (azimuthal direction of $\vec{q_{T}}$) is about 0.3--0.45 radians depending on the jet energy. We use a Monte Carlo method to estimate the resulting ``dilution factors'' due to smearing on the amplitude of the sine modulation. We find multiplicative factors of about 1.03, which is negligible for the purposes of this study.

The resolution of the Collins angle (azimuthal direction of $\vec{j_{T}}$) is driven by the interplay between the hadron momentum and jet-energy resolutions; however, the jet-energy resolution always dominates for EIC energies (for the tracking resolution shown in Table~\ref{tab:resolutions}). Depending on the $z_h$, the relative resolution on the Collins angle ranges from 0.06 to 0.25 rad for $20<E ^{\mathrm{jet}}<30$ GeV, from 0.05 to 0.20 rad for $30<E^{\mathrm{jet}}<40$ and from 0.05 to 0.10 rad for $40<E^{\mathrm{jet}}<50$.  These resolutions compare favorably to the performance achieved in the hadron-in-jet measurements by STAR in both the charged-pion channel~\cite{Adkins:2017iys} and neutral-pion channel~\cite{Pan:2014bea}. We find that the associated ``dilution factors'' are negligible. 
\subsection{Particle ID requirements}
\label{sec:pid}

The hadron-in-jet measurement requires particle identification (PID) to provide the flavor sensitivity that is critical for the interpretation of the data in terms of the Collins FF and quark transversity. While \textsc{Delphes} does have the capability of emulating PID detectors, we do not use that feature as estimates for a momentum-dependent performance are not yet available. Instead, we perform a study that illuminates the PID requirements for the studies presented in Section~\ref{sec:hadroninjet}.

\begin{figure}
    \centering
    \includegraphics[width=0.49\textwidth]{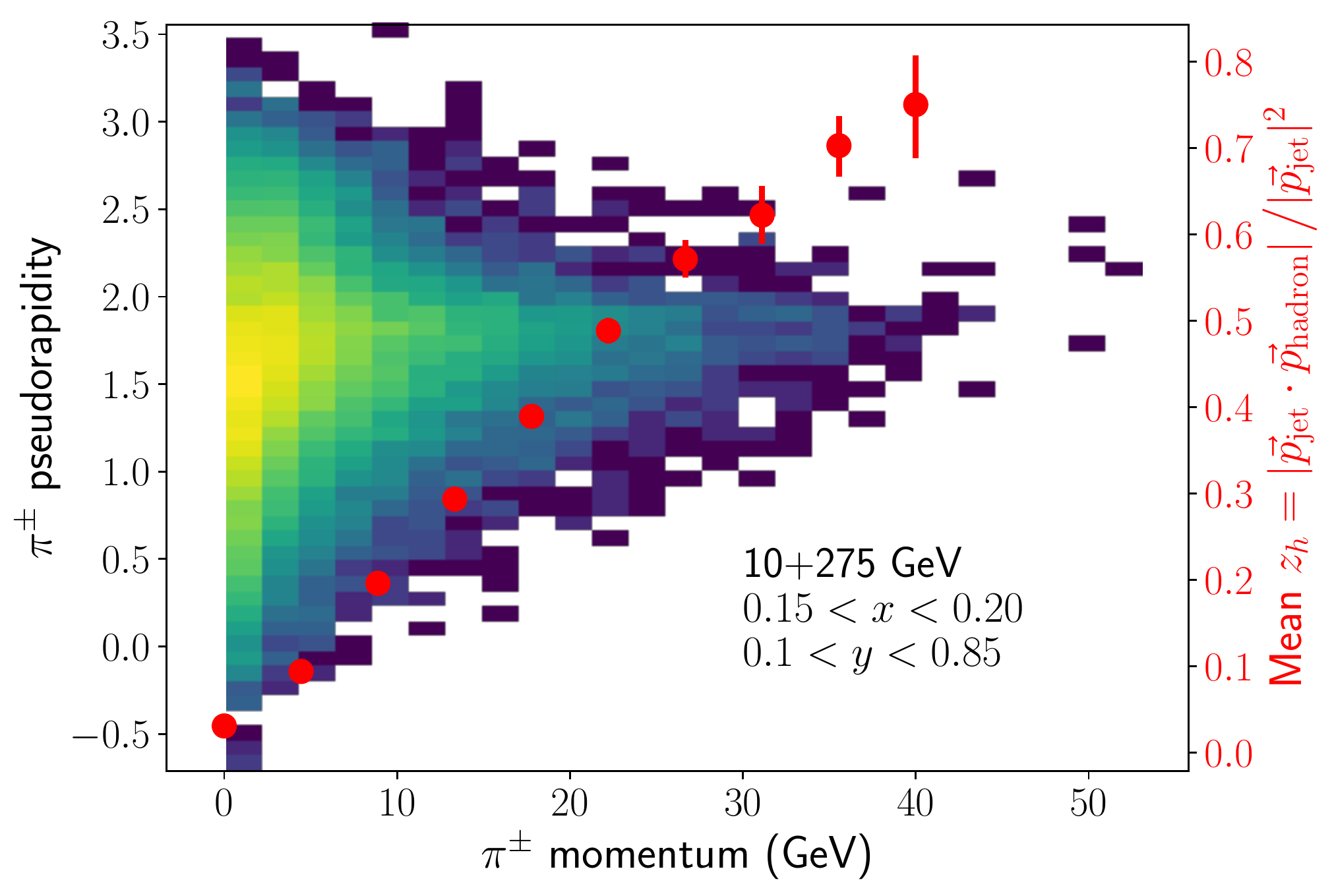}
\caption{\label{fig:PIDrequirement}
Pseudorapidity and momentum distribution for charged-pions in jets with $p_{T}>5$~GeV. The average longitudinal momentum fraction of the hadron with respect to the jet axis is shown by the red dots.}
\end{figure}

Figure~\ref{fig:PIDrequirement} shows the momentum and pseudorapdity distribution of charged pions in jets for events with $0.1<x<0.2$, as well as the average $z_h$ value sampled in each momentum interval. Positive particle identification of pions up to $\approx$ 40 GeV at $\eta\approx$ 1.5--2.0 is required to reach $z_h\approx 0.8$. Smaller $x$ ranges yield smaller jet momentum and thus less stringent requirements. 

\subsection{Systematic uncertainties}
\label{sec:systematics}

Most sources of systematic uncertainties in jet measurements, including jet-energy scale (JES) and jet-energy resolution (JER) uncertainties, cancel in the spin-asymmetry ratios. Time drifts in the detectors response can be suppressed to a negligible level with the bunch-to-bunch control of the beam polarization pioneered at RHIC, which will transfer to EIC. 

While the JES uncertainty does not affect the scale of the asymmetry, it affects the definition of $z_h$ (the jet momentum  appears in the denominator) or $q_{T}$ (proportional to jet momentum), so it translates to a horizontal uncertainty in the differential asymmetry measurements. We show that a conservative estimate of 3$\%$ for the JES uncertainty would still allow us to sample the predicted $z_{h}$-dependence of the Collins asymmetries shown in Figure~\ref{fig:Collins} or the Sivers asymmetry shown in Figure~\ref{fig:Sivers}. 

While the asymmetry measurements have the potential to be very accurate, the unpolarized cross section measurement will be much more challenging due to the JES uncertainty. HERA experiments ultimately achieved a JES uncertainty of about 1\%~\cite{Newman:2013ada}, but there are several challenges for the EIC. The accelerator design that leads to an improvement of the instantaneous luminosity compared to HERA requires focusing magnets closer to the interaction point. This limits the space for detectors, which will result in ``thin'' hadronic calorimeters that motivate the constant terms in Table~\ref{tab:resolutions}; this will also lead to more difficult JES estimates. 

While difficult, the measurement of the unpolarized cross sections is crucial to constrain nonperturbative aspects of TMD-evolution which is not only motivated by the need to understand the hadronization process itself but ultimately improves the accuracy of the extractions of the Sivers function~\cite{Echevarria:2014xaa}. 

Estimations of the jet energy scale uncertainty are notoriously difficult and involve several studies that cover beam-test data, in-situ calibrations, and Monte-Carlo simulations (e.g. Ref.~\cite{Khachatryan:2016kdb}), which are outside the scope of this work. 

Systematic uncertainties that do not cancel in the asymmetry ratio are the ones associated with the relative luminosity for each polarization state and the beam polarization. The relative luminosity uncertainty will be $<0.1\%$, as demonstrated at RHIC. The relative uncertainty on the hadron polarization is expected to be $<1\%$ at the EIC. Given the absolute magnitude of the  Sivers and Collins asymmetries we predict, neither of these uncertainties will be a limiting factor. 

The systematic uncertainties associated with the underlying event, which were dominant at low \pTjet~in Sivers- and Collins-asymmetry studies at RHIC~\cite{Abelev:2007ii,Adamczyk:2017wld}, will be negligible given that high $Q^{2}$ DIS is essentially free from ambiguities due to the beam-remnant (as illustrated in Figure~\ref{fig:polarplot}).

\section{Conclusions~\label{sec:conclusions}}

We have presented predictions and projections for measurements of the Sivers asymmetry with electron-jet azimuthal correlations and the Collins asymmetry with hadron-in-jet measurements at the EIC. In particular, we have presented for the first time predictions for Collins asymmetries using hadrons inside jets and we argued that it will be a key channel to access quark transversity, Collins fragmentation functions, and to study their evolution. 

We have explored the feasibility of these measurements based on fast simulations implemented with the \textsc{Delphes} package and found that the expected performance of a hermetic EIC detector with reasonable parameters is sufficient to perform these measurements. We have discussed detector requirements and suggested further studies to go along with dedicated detector simulations to inform the design of future EIC experiments, which we argue should include jet capabilities from day one.  

While jet-based measurements of Sivers and transversity functions are powerful and novel ways to achieve some of the main scientific goals of the EIC, the potential of jets transcends these two examples. A promising case are novel jet substructure studies for TMD observables, which we leave for future work. This work represents a new direction for the rapidly emerging field of jet studies at the future EIC~\cite{Arratia:2020azl,Arratia:2020ssx,Borsa:2020ulb,Peccini:2020tpj,Guzey:2020gkk,Guzey:2020zza,Kang:2020xyq,Arratia:2019vju,Page:2019gbf,Li:2020sru,Gutierrez-Reyes:2019vbx,Gutierrez-Reyes:2019msa,Zhang:2019toi,Aschenauer:2019uex,Hatta:2019ixj,Mantysaari:2019csc,DAlesio:2019qpk,Kishore:2019fzb,Kang:2019bpl,Roy:2019cux,Roy:2019hwr,Salazar:2019ncp,Gutierrez-Reyes:2018qez,Boughezal:2018azh,Klasen:2018gtb,Dumitru:2018kuw,Liu:2018trl,Zheng:2018ssm,Sievert:2018imd,Klasen:2017kwb,Hinderer:2017ntk,Chu:2017mnm,Aschenauer:2017jsk,Abelof:2016pby,Hatta:2016dxp,Dumitru:2016jku,Boer:2016fqd,Dumitru:2015gaa,Hinderer:2015hra,Altinoluk:2015dpi,Kang:2013nha,Pisano:2013cya,Kang:2012zr,Kang:2011jw,Boer:2010zf}.

\section*{Code availability}
The Delphes configuration file for the EIC general-purpose detector considered in this work can be found at: 
\url{https://github.com/miguelignacio/delphes_EIC/blob/master/delphes_card_EIC.tcl}{} 

\section*{Acknowledgements} \label{sec:acknowledgements}
We thank Oleg Tsai for insightful discussions on calorimetry technology for EIC detectors. We thank Elke Aschenauer, Barbara Jacak, Kyle Lee, Brian Page and Feng Yuan for enlightening discussions about jet physics at the EIC and Anselm Vossen, Feng Yuan, Sean Preins, and Sebouh Paul for feedback on our manuscript. We thank the members of the EIC User Group for many  insightful discussions during the Yellow Report activities. M.A and A.P. acknowledges support through DOE Contract No. DE-AC05-06OR23177 under which Jefferson Science Associates, LLC operates the Thomas Jefferson National Accelerator Facility. Z.K. is supported by the National Science Foundation under Grant No.~PHY-1720486 and CAREER award~PHY-1945471. A.P. is supported by the National Science Foundation under Grant No.~PHY-2012002. F.R. is supported by LDRD funding from Berkeley Lab provided by the U.S. Department of Energy under Contract No.~DE-AC02-05CH11231 as well as the National Science Foundation under Grant No.~ACI-1550228.

\appendix

\section{Relevant perturbative results at one-loop~\label{app:one}}

Here we summarize the different functions that appear in the factorization formulas in Eqs.~(\ref{eq:crosssection1}) and~(\ref{eq:crosssection2}). We work in Fourier-transform space where all the associated renormalization group equations are multiplicative. They can be derived from the fixed-order result along with the relevant anomalous dimensions. In the unpolarized case we have
\begin{align}
    H_q(Q,\mu)=
    &\,
    1+\frac{\alpha_s}{2\pi}C_F\bigg[-\ln^2\Big(\frac{\mu^2}{Q^2} \Big)-3\ln\Big(\frac{\mu^2}{Q^2}\Big)
    \nn\\&\,
    -8+\frac{\pi^2}{6}\bigg]
    \,,\\
    J_q(p_TR,\mu)=
    &\,
    1+\frac{\alpha_s}{2\pi}C_F\bigg[\frac12 \ln^2\Big(\frac{\mu^2}{p_T^{2}R^2}\Big)
    \nn\\&
    +\frac32\ln\Big(\frac{\mu^2}{p_T^{2} R^2}\Big)
    +\frac{13}{2}-\frac{3}{4}\pi^2\bigg]
    \,,\\
    S_{q}(\vec b_T,y_{\rm jet},R,\mu)=&\, 1 + \frac{\alpha_s}{2\pi}C_F\bigg[-\ln\Big(\frac{e^{-2y_{\rm jet}}}{R^2}\Big)\ln\Big(\frac{\mu^2}{\mu_b^2}\Big)
    \nn\\&\,
    -\frac12 \ln^2\Big(\frac{1}{R^2}\Big)\bigg]\,,
\end{align}
where $\mu_b=2e^{-\gamma_E}/b_T$. The factorization here holds for $R\sim{\cal O}(1)$. Note that all $\ln\mu$ terms cancel at fixed order.

The unpolarized TMD PDF and FF can be matched onto the collinear PDFs at low values of $b_T $ as
\begin{align}
f_q^{\rm TMD}(x_B,\vec b_T,\mu_b)&\simeq  
\sum_i 
\int_{x_B}^1 \frac{dx}{x} \,
C_{q\gets i} \left(\frac{x_B}{x}, \mu_b \right) f_{1}^{i}(x,\mu_b),\label{convx}
\\
D_{h/q}^{\rm TMD}(z_h,\vec b_T,\mu_b) &\simeq 
\sum_j
 \int_{z_h}^1 \frac{d z}{z} \,
\hat {C}_{j\gets q}\left(\frac{z_h}{z}, \mu_b \right) D_{h/j}(z,\mu_b).\label{convz}
\end{align}
where according to Ref.~\cite{Meng:1995yn,Nadolsky:1999kb,Koike:2006fn,Kang:2015msa},
\begin{align}
C_{q\gets q'}(x,\mu_b) = &\delta_{q'q} \Big[\delta(1-x)   \nonumber\\
&   + \frac{\alpha_s}{\pi}\left(\frac{C_F}{2}(1-x)  \right)\Big]\; , \label{eq:cf_css}
\\
C_{q\gets g}(x,\mu_b) = & \frac{\alpha_s}{\pi} {T_R} \, x (1-x)\; ,  \label{eq:cf1_css}
\\
\hat C_{q'\gets q}(z,\mu_b) = &\delta_{q'q} \Big[\delta(1-z) + \frac{\alpha_s}{\pi}\Big(\frac{C_F}{2}(1-z)\nonumber \\
&   + P_{q\gets q}(z)\, \ln z\Big) \Big]\; ,  \label{eq:cd_css}
\\
\hat C_{g\gets q}(z,\mu_b) = & \frac{\alpha_s}{\pi} \left( \frac{C_F}{2} z\; +  P_{g\gets q}(z)\, \ln z \right) .  \label{eq:cd1_css}
\end{align}
with the usual splitting functions $P_{q\gets q}$ and $P_{g\gets q}$ given by
\begin{align}
P_{q\gets q}(z) &= C_F \left[ \frac{1+z^2}{(1-z)_+} + \frac{3}{2} \delta(1-z) \right] \, , 
\label{P_qq}
\\
P_{g\gets q}(z) &= C_F \frac{1+(1-z)^2}{z} \; .
\label{P_gq}
\end{align}

The  energy evolution of TMDs from the scale $\mu_b$ to the scale $Q$ 
is encoded in the exponential factor, $\exp[-S_{\rm sud}]$, with the Sudakov-like form factor, the perturbative 
 part of which can be written as
 \begin{align}
S_{\rm pert}(Q,b)=\int_{\mu_b^2}^{Q^2}\frac{d\bar\mu^2}{\bar\mu^2}\left[A(\alpha_s(\bar\mu))\ln\frac{Q^2}{\bar\mu^2}+B(\alpha_s(\bar\mu))\right] \, .
\label{spert}
\end{align}
Here the coefficients $A$ and $B$ can be expanded as a
perturbative series $A=\sum_{n=1}^\infty A^{(n)} \left(\alpha_s/\pi\right)^n$, $B=\sum_{n=1}^\infty B^{(n)} \left(\alpha_s/\pi\right)^n$.
In our calculations,
we take into account $A^{(1)}$, $A^{(2)}$ and $B^{(1)}$ to achieve NLL accuracy. Because this part is spin-independent, these coefficients are the same for the polarized and unpolarized cross sections~\cite{Collins:1984kg} and are given by~\cite{Kang:2011mr,Aybat:2011zv,Echevarria:2012pw,Collins:1984kg,Qiu:2000ga,Landry:2002ix}:
\begin{align}
A^{(1)} &=C_F\,,
\; \nonumber \\
A^{(2)}
&=\frac{C_F}{2}\left[C_A\left(\frac{67}{18}-\frac{\pi^2}{6}\right)-\frac{10}{9}T_R \, n_f\right]\,,
\; \nonumber \\
B^{(1)} &= -\frac{3}{2} C_F
\,. 
\end{align}
In order to avoid the Landau pole $\alpha_s(\mu_b)$, we use the standard $b_*$-prescription that introduces a cutoff value $b_{\rm max}$ and allows for a smooth transition from the perturbative to the nonperturbative region,  
\begin{equation}
b_T \Rightarrow b_*=\frac{b_T}{\sqrt{1+b_T^2/b_{\rm max}^2}}  \,,
\end{equation}
where $b_{\rm max}$ is a parameter of the prescription. From the above definition, $b_*$ is always in
the perturbative region where $b_{\rm max}$ was chosen~\cite{Kang:2015msa} to be 1.5~GeV$^{-1}$. When $b_*$ is introduced in the Sudakov form factor, the total Sudakov-like form factor can be written as the sumof the perturbatively calculable part and a nonperturbative contribution
\begin{equation}
{S}_{\rm sud}(Q;b_T)\Rightarrow { S}_{\rm pert}(Q;b_*)+S_{\rm NP}(Q;b_T) \, ,
\end{equation}
where $S_{\rm NP}(Q;b_T) = S_{\rm NP}^{f}(Q;b_T) + S_{\rm NP}^{D}(Q;b_T)$ is defined as the
difference
between the original form factor and the perturbative one. Eventually, we have
\begin{align}
f_q^{\rm TMD}(x_B,\vec b_T, Q)&=  f_q^{\rm TMD}(x_B,\vec b_T,\mu_b) \nn \\
&\times e^{-\frac{1}{2}{ S}_{\rm pert}(Q;b_*)-S_{\rm NP}^{f}(Q;b_T)}
\\
D_{h/q}^{\rm TMD}(z_h,\vec b_T, Q) &= 
D_{h/q}^{\rm TMD}(z_h,\vec b_T,\mu_b)\nn \\
&\times e^{-\frac{1}{2}{ S}_{\rm pert}(Q;b_*)-S_{\rm NP}^{D}(Q;b_T)}
\end{align}
In our calculations we use the prescriptions for the nonperturbative functions and the treatment of the Collins fragmentation and Sivers functions of Refs.~\cite{Kang:2015msa,Echevarria:2014xaa}.

\FloatBarrier
\renewcommand\refname{Bibliography}
\bibliographystyle{utphys} 
\bibliography{bibio.bib} 

\appendix*

\end{document}